\documentstyle[12pt]{article}
\begin{document}
\title{
On Multistep Bose-Einstein Condensation \\ in Anisotropic Traps}
\author{K. Shiokawa\thanks
{  E-mail address: kshiok@phys.ualberta.ca }
\\
{\small Department of Physics, University of Maryland, College Park,
MD 20742, USA}\\
\small and \\
{\small Center for Nonlinear Studies, Hong Kong Baptist University,
Kowloon Tong, Hong Kong
}\\
\small and \\
{\small Theoretical Physics Institute,
University of Alberta,
Edmonton, Alberta T6G 2J1, Canada
}\\
\small{\it(Umdpp 99-076, Alberta Thy 22-99, to appear in Jour. of Phys. A)}
}
\maketitle
\begin{abstract}
 Multistep Bose-Einstein condensation
 of an ideal Bose gas in anisotropic harmonic atom traps is studied.
 In the presence of strong anisotropy realized by the different trap frequency
 in each direction,
 finite size effect dictates a series of dimensional crossovers
 into lower-dimensional excitations.
 Two-step condensation and
 the dynamical reduction of the effective dimension
 can appear in three separate steps.
 When the multistep behavior occurs,
 the occupation number of atoms excited in each dimension is shown to behave
 similarly as a function of the temperature.
% Depending on the density of states of the system,
Multistep behaviors
%similar results are expected in systems with finite geometry
can be easily controlled by changing the degree of anisotropy.
\end{abstract}
\input epsf
\section{Introduction}
%
%%%%%%%%%%%%%%%%%%%%%%%%%%%%%%%%%%%%%%%%%%%%%%%%%%%%%%%%%%%%%
%      Introduction
%%%%%%%%%%%%%%%%%%%%%%%%%%%%%%%%%%%%%%%%%%%%%%%%%%%%%%%%%%%%%
Although Bose-Einstein condensation (BEC) has been an active topic
of research in condensed matter physics for decades
\cite{Bose24,Einstein24,Huang,Pathria}, recent revival of
interests in this field is mainly credited to the achievement in
atom molecular optics; most recent laser cooling and evaporative
cooling techniques in magnetic and optical atom traps enable us to
realize BEC of a weakly interacting gas in a controlled way
\cite{AEMWC95,DMADDKK95,MADKDK96}. The weakly interacting nature
of a gas allows us to handle the phenomena and make theoretical
predictions with high accuracy. The rapid progress in this field
also stimulates other areas in physics such as high energy physics
and astrophysics \cite{BEC95}.
%It also provides the nice testing ground for fundamental aspects of
%quantum mechanics.

%%%%%%%%%%%%%%%%%%%%%%%%%%%%%%%%%%%%%%%%%%%%%%%%%%%%%%%%%%%%%
%      Finite Size Effect in BEC
%%%%%%%%%%%%%%%%%%%%%%%%%%%%%%%%%%%%%%%%%%%%%%%%%%%%%%%%%%%%%
For atom trap experiments, since atoms are trapped in a finite
geometry, finite size as well as finite number effects play a
significant role in the condensation process. The conventional
phase transition picture defined in the thermodynamic limit has to
be reexamined or modified
\cite{Krueger68,Sonin69,Osborne49,BagKle91,GroHol95b,
KirTom96,GroHol95c,IngoldLambrecht98}.
One significant change due to finiteness
is the existence of BEC in low-dimensional systems.
%%%%%%%%%%%%%%%%%%%%%%%%%%%%%%%%%%%%%%%%%%%%%%%%%%
% add
%%%%%%%%%%%%%%%%%%%%%%%%%%%%%%%%%%%%%%%%%%%%%%%%%%%%
Recently,
quasi low-dimensional
systems prepared by optical or magnetic trapping devices
are actively discussed \cite{GHSSPM98,VKCC99}.
The study of such systems provides an ideal test for the theory of
finite size, low-dimensional systems in a controlled environment.
%%%%%%%%%%%%%%%%%%%%%%%%%%%%%%%%%%%%%%%%%%%%%%%%%%
% add
%%%%%%%%%%%%%%%%%%%%%%%%%%%%%%%%%%%%%%%%%%%%%%%%%%%%

The critical behavior of a finite size system
\cite{Fisher71,BarFis73,Barber83} is characterized by the
effective infrared dimension (EIRD) of the system
\cite{HuOco84,OSH88}:
 near the critical point when the contribution of
the lowest mode of a system dominates, its symmetry properties
can be shown to be equivalent to a lower-dimensional one. The
system, in such a case, is said to possess an EIRD. Varying the
relative size (or shape) of the potential changes the infrared
behavior and hence the EIRD of the system.

Dimensionless parameters
$\eta_i = \beta \hbar \omega_i$ ($\beta= 1 / k_{B} T$)
for a harmonic oscillator potential with natural
frequencies $\omega_i ~(i = 1,2,3)$
characterize the degree of anisotropy and finite size effects.
With respect to $\eta_i$, we can classify the dynamical behavior of the %%@
system
into the following four cases
dependent on the degree of anisotropy:

Case 1; ~~$\eta_1, \eta_2, \eta_3 > 1$
$~~\rightarrow$ EIRD
 = 0,

Case 2; ~~$\eta_1, \eta_2 > 1 > \eta_3$
$\rightarrow$ EIRD = 1,

Case 3; ~~$\eta_1 > 1 > \eta_2, \eta_3$
$\rightarrow$ EIRD = 2,

Case 4; ~~$1 > \eta_1, \eta_2, \eta_3$
$~~\rightarrow$ EIRD = 3.  \\
\noindent
As the temperature is lowered,
dynamical dimensional reduction of the system
characterized by the decrease of effective dimension
can be observed.
Particularly, in the presence of maximal anisotropy
$ 1 >> \eta_1 >> \eta_2 >> \eta_3$, EIRD decreases one by one
from three to zero at the temperature
$k_{B} T = \hbar \omega_1, \hbar \omega_2$, and $\hbar \omega_3$.

While the picture described above is of generic nature of any
quantum system, for particles that obey Bose statistics, the
dimension of the dynamics of excited particles can also change in
the form of condensation.
%Depending on the density of states of the system,
BEC can also occur in separate steps in the presence of anisotropy
reducing the effective dimension attributed to the excited modes
of the system by one, two, or
three at a time \cite{Sonin69}.
In this sense,
the dimensional crossover
associated with both the multistep condensation and the reduction of
effective dimension due to frozen degrees of freedom
can manifest themselves in a similar way in a finite system
in spite of their different origin.
This is the main subject discussed in this paper.

For the system with finite size and number
of atoms, the reduced chemical potential $\phi \equiv \beta (E_0 -
\mu)$ does not strictly vanish at the critical temperature. Only
in the thermodynamic limit, $\phi$ vanishes at the critical
temperature and the specific heat develops the discontinuity in
the derivative at the critical point \cite{BarFis73}. In an
isotropic system, the thermodynamic limit can be uniquely defined
as discussed in many textbooks \cite{Pathria}.
 However, if we allow anisotropy in the system,
 there are three different ways of taking a thermodynamic limit:
(Three-dimensional limit) $\omega_1, \omega_2, \omega_3
\rightarrow 0$ while keeping $N \omega_1 \omega_2 \omega_3$ fixed.
In this case,
the system shows the critical behavior of three dimension and the
corresponding three dimensional critical temperature $T_{3D}$ can
be defined.
(Two-dimensional limit) $\omega_2 \omega_3 \rightarrow 0$ while $N
\omega_2 \omega_3$ fixed.
 The system shows the critical behavior
of two dimension and the corresponding critical temperature
$T_{2D}$ is the one for the two-dimensional system.
(One-dimensional limit) If we simply take $\omega_3 \rightarrow 0$
while $N \omega_3$ fixed, the critical temperature vanishes.
However, if we tune $\omega_3$ a little slower such that $\omega_3
\sim \log(2N) / N \rightarrow 0$, the one-dimensional critical
temperature $T_{1D}$ can still be defined in this modified sense.
We will discuss this issue again in Sec. 3.1.

In the presence of strong anisotropy, the whole particle spectrum
naturally splits into zero, one, two, and three-dimensional
excitations. The ground state is viewed as a zero-dimensional
excitation. Let us denote the number of modes excited in the
corresponding directions as $N_0$, $N_1$, $N_2$, and $N_3$,
respectively. An $n$-dimensional condensation temperature $T_{nD}$
($n=1,2,3$) can be defined as the temperature at which all the
$n$-dimensionally excited modes are saturated:
\begin{eqnarray}
\mbox{3-dimensional;} ~N&=&N_3(T_{3D}),
  \label{def3d}     \\
\mbox{2-dimensional;} ~N&=&N_3(T_{2D}) + N_2(T_{2D}),
  \label{def2d}     \\
\mbox{1-dimensional;}
~N&=&N_3(T_{1D}) + N_2(T_{1D}) + N_1(T_{1D}).
  \label{def1d}
\end{eqnarray}
%
%If the thermodynamic limit is taken differently in each case as in (a),(b), %%@
%and (c),
One can see that the condensation temperatures defined above are
equivalent to the critical temperatures if the appropriate
$n$-dimensional thermodynamic limit is taken. This splitting of
the excitation spectrum gives the basis of the rest of our analysis.
Similar splitting was proposed in \cite{Sonin69}
for liquid helium.
In a liquid, however, an occupation number of particles excited
in a particular direction is extremely difficult to observe.
Furthermore, the validity of this splitting is by no
means obvious for the strongly interacting system such as the
liquid helium. On the other hand, such a quantity is directly
observable in atom trap experiments
and therefore it deserves careful study.
Moreover, as shown in
Section 3, occupation numbers $N_1$, $N_2$, and $N_3$ behave
as if they were independent quantities and show
the similar behavior when the multistep crossover occurs.
This result indicates the independent nature of each $N_i$
for strongly anisotropic systems.
%$T_{nD}$ can be obtained by tuning n out of
%three anisotropy parameters $\eta_i$
%to reach zero
%in addition to $N \rightarrow \infty$.
%Anisotropy parameters $\eta_1 / \eta_3$
%and $\eta_2 / \eta_3$ are varied accordingly.
%%Finite size corrections
%necessarily modify above definitions, since they involve
%the excitations in lower dimensions.
%In Section 3, we will discuss this aspect in detail.

For a realistic system where $N$, $\omega_1, \omega_2, \omega_3$,
are all finite and fixed, physically observable temperature of
interest is the crossover temperature
%$T_{nD}^{*}$ ($n=1,2,3$)
at which the deviation from the bulk critical behavior sets in.
%The evaluation of the crossover temperature requires us to take
%the finite size effects into account,
%accordingly, we must modify above definitions, since each $T_{nD}^{*}$
%contains the information of lower dimensional excitations
%as higher order terms.
The crossover temperature is achieved when the correlation length
reaches the size of the system since further ordering in this
direction will be suppressed at this point.
In the strongly
anisotropic system in which $\omega_1, \omega_2 >> \omega_3$
holds, $T_{1D}, T_{2D} << T_{3D}$ gives the necessary
(but not sufficient) condition for the multistep condensation:
the condensation into two, one-dimensional modes and into
the ground state can occur in separate steps.
The multistep condensation was discussed
for a nonrelativistic ideal gas in a cavity \cite{Sonin69}, in a
harmonic trap \cite{DruKet97}, and a relativistic ideal gas in a cavity
\cite{HuShi98}.

In Section 2, we study the excitation spectrum of anisotropic
harmonic oscillators. We focus our attention to three different
cases in which the equipotential surface has a prolate, oblate,
and maximally anisotropic ellipsoidal shape. In Section 3, we show
that each case shows qualitatively different condensation
behavior. After introducing the condensation temperatures defined
in the bulk limit, we focus on the multistep crossover behavior of
excited modes between different effective dimensions through BEC
or dynamical reduction of EIRD.
 In particular, in a maximally anisotropic potential,
 the dimensional reduction can occur in three-steps.
 In such a case, we show that
each dimensional component behaves in a similar manner as a
function of the reduced temperature defined near the corresponding
crossover temperature.
The possible effect of interactions is also discussed.

This work deals with a nonrelativistic ideal Bose gas in
anisotropic magnetic traps, and a companion paper deals with a
relativistic ideal Bose gas in rectangular cavities
\cite{HuShi98}. Our calculations are focused on the occupation
number and condensation temperature for each dimension. The effect
of an interaction in condensation process has been discussed in
many literatures. In principle,
it can affect the dynamics of condensation considerably.
For a weakly interacting gas, however, the averaged quantities
such as condensation fractions and critical temperatures are
relatively insensitive to the presence of interactions and the
corrections to bulk ideal-gas value
 are well-explained by the
finite number correction \cite{BPK87,EJMWC96}. On the other hand,
interaction effects are known to affect higher moments such as
specific heat significantly and considered to be essential to
explain the observed specific heat data. Throughout the rest of
the paper, we use units such that $k_B = \hbar = 1$ for brevity.
The results in ordinary units can be easily reproduced by
replacing $\omega \rightarrow \hbar \omega$ and $T \rightarrow k_B
T$.

%
%
%            Dynamical Symmetry
%

\section{Anisotropic Harmonic Oscillator and Excitation Spectrum}

For an anisotropic harmonic oscillator
with oscillator frequencies
$\omega_{i}~(i=1,2,3)$, the Hamiltonian has the form
\begin{equation}
    H = \frac{1}{2} \sum_{i=1}^{3} (p_i^2 + \omega_i^2 x_i^2) .
      \label{f0}
\end{equation}
In this paper, we study cases where
the frequencies $\omega_{i}$ are rationally related.\footnote{
This assumption of rationality is rather for the technical convenience
and will not affect the physical results of this paper.
Other methods can be found, for example, in \cite{KirTom96}, \cite{DruKet97},
and \cite{HHR97}.}
Whence there exist integers $k_{i} ~(i=1,2,3)$ such that
$\omega_{i} k_{i} = \omega ~(i=1,2,3)$ where
$ \omega  = \Omega (k_1 k_2 k_3)^{1/3}$ and
$ \omega_1 \omega_2 \omega_3 = \Omega^3$
.

The energy level of an anisotropic harmonic oscillator is given by
\begin{equation}
    E_{n} = \sum_{i=1}^{3} \omega_{i} (n_i + 1/2)  .
      \label{f1}
\end{equation}
We can also define the energy level modulo $k_i$ as
$ n_i = k_i \nu_i + \lambda_i $
where $ \nu_i \equiv [ n_i / k_i ]$,
and $[~]$ denotes the integer part of the number inside the bracket.
Then Eq. (\ref{f1}) can be written as
\begin{equation}
    E_{n} = \omega M +
    \sum_{i=1}^{3} \omega_i \lambda_i + E_0,
      \label{f2}
\end{equation}
where $M = \nu_1 + \nu_2 + \nu_3$.
The first term in Eq. (\ref{f2}) corresponds to the isotropic
harmonic oscillator Hamiltonian (see also Appendix A).
The ground state energy $E_0$ has the familiar form
\begin{equation}
    E_0
    = \frac{ \omega_1  + \omega_2  + \omega_3  }{ 2 } .
      \label{c3f7}
\end{equation}

\subsection{Excitation spectrum}

%%%%%%%%%%%%%%%%%%%%%%%
%
%  1d condensation
%
%%%%%%%%%%%%%%%%%%%%%%%
%\subsection{One-Dimensional Condensation}

%%%%%%%%%%%%%%%%%%%%%%%
%
%  1d condensation
%
%%%%%%%%%%%%%%%%%%%%%%%
\subsubsection{Prolate shape potential}

First we discuss the case of anisotropy
corresponding to
$\omega_1 = \omega_2 > \omega_3$
(we simply choose here $k_1 = k_2 = 1 < k_3$).
In such a case,
the equipotential surface has a prolate shape.
For a strong anisotropy $k_3 >> 1$,
two-step condensation can occur.
In such a case, $\omega = \omega_1 = \omega_2$ and the energy eigenvalue is
\begin{equation}
    E_n =
    \omega  M
    + \omega_3 \lambda_3 + E_0      .
      \label{c3f}
\end{equation}

For sufficiently large values of $k_3$,
the whole energy spectrum can be split into
the energy level of
the ground state ( $\bar{E}_{n} = 0$),
  one-dimensionally excited states
 ( $\bar{E}_{n} = n_3 \omega_3 $; $n_3 = 1, 2, \cdots$),
 two-dimensionally excited states
 ( $\bar{E}_{n} = n_1 \omega_1 + n_3 \omega_3 $;
$n_1 = 1, 2, \cdots, n_3 = 0,1,\cdots$
  and $n_2 \omega_2 + n_3 \omega_3 $,
$n_2 = 1, 2, \cdots, n_3 = 0,1,\cdots$),
  and three-dimensionally excited states
 ( $\bar{E}_{n} = n_1 \omega_1 + n_2 \omega_2 + n_3 \omega_3 $;
  $n_1, n_2 = 1, 2, \cdots, n_3 = 0, 1, \cdots$ ),
where $\bar{E}_n \equiv E_n - E_0 =
    \omega  M + \omega_3  \lambda_3  $ is the energy measured from the
    ground state.
 The number of particles excited
 in these dimensions are given respectively by
\begin{eqnarray}
  N_0 &=& \frac{z}{1-z},
    \\        \label{f90}
  N_1 &=& \sum_{n_3=1}^{\infty} \frac{z}{e^{ n_3 \eta_3 } - z},
            \label{f91}  \\
  N_2 &=& \sum_{n_1=1,n_3=0}^{\infty}
  \frac{2 z}{e^{ n_1 \eta_1 + n_3 \eta_3 } - z}
    \nonumber  \\
        &=& \sum_{\lambda_3=0}^{k_3 - 1}  \sum_{M=1}^{\infty}
  \frac{2 M z}{e^{ M \eta_1 + \lambda_3 \eta_3 } - z},
              \label{f92} \\
  N_3 &=& \sum_{n_1=n_2=1,n_3=0}^{\infty}
  \frac{z}{ e^{ n_1 \eta_1 + n_2 \eta_2 + n_3 \eta_3 } - z}
    \nonumber  \\
        &=& \sum_{\lambda_3=0}^{k_3 - 1}  \sum_{M=2}^{\infty}
  \frac{(M-1)M}{2}
  \frac{z}{ e^{ M  \eta_1 + \lambda_3 \eta_3 } - z},
                    \label{f93}
\end{eqnarray}
where $z \equiv e^{\beta(\mu - E_0)}$ is the (reduced) fugacity.
The factor $2$ in Eq. (\ref{f92}) accounts for the symmetry between
the first and second axis.
These expressions can be further simplified in the following manner.

%%%%%%%%%%%%%%%%%%%%%%%
%
% N1
%
%%%%%%%%%%%%%%%%%%%%%%%
For one dimension,
\begin{eqnarray}
  N_1 &=&
    \sum_{n_3=1}^{\infty}
    \frac{ z e^{ - n_3 \eta_3 } }{ 1 - z e^{ - n_3 \eta_3 } }
    =
    \sum_{l=1}^{\infty}
    \frac{ z^l e^{ - l \eta_3 } }{ 1 - e^{ - l \eta_3 } }
    \\
      &=&
    \frac{ g_1(z e^{ - \eta_3 / 2 } ) }{ \eta_3 } + \cdots,
    \label{f10}
\end{eqnarray}
where $ g_p(z) = \sum_{ n=1}^{\infty} \frac{z^n}{n^p}   $
is the Bose-Einstein function.
We used Eq. (\ref{App0}) to obtain the second line from the first line.
%%%%%%%%%%%%%%%%%%%%%%%
%
% N2
%
%%%%%%%%%%%%%%%%%%%%%%%

For two-dimensional excitations,
making use of Eq. (\ref{App2}),
Eq. (\ref{f92}) can be written as
\begin{eqnarray}
    N_2
     &=&
         2 \sum_{\lambda_3=0}^{k_3 - 1}  \sum_{M=1}^{\infty}
         \sum_{l=1}^{\infty} M z^l e^{-l (M \eta_1 + \lambda_3 \eta_3) }
     \nonumber  \\
     &=&
        2 \sum_{l=1}^{\infty}
        \frac{ z^l e^{-l \eta_1}}
        { ( 1 - e^{-l \eta_1} ) ( 1 - e^{-l \eta_3} ) }
     \nonumber  \\
     &=&
        2 \sum_{l=1}^{\infty}
         \frac{ z^l e^{- l (\eta_1 - \eta_3) / 2 }}
        { l^2 \eta_1 \eta_3 }
    -
                ( k_3 + \frac{1}{k_3} )
               \sum_{l=1}^{\infty}
               \frac{ z^l e^{   - l (\eta_1 - \eta_3) / 2 }}
                { 12 }
         + \cdots
     \nonumber  \\
     &=&
    \frac{ 2 g_2( z e^{- (\eta_1 - \eta_3) / 2 }  )}{ \eta_1 \eta_3 }
        -
      \frac{ k_3 g_0( z e^{- (\eta_1 - \eta_3) / 2 }  )}{ 12 }
         + \cdots.
        \label{f15}
\end{eqnarray}
To obtain the third line from the second line, we used Eq. (\ref{App3}).

%%%%%%%%%%%%%%%%%%%%%%%
%
% N3
%
%%%%%%%%%%%%%%%%%%%%%%%

For three dimensional excitations, Eq. (\ref{f93}) gives
\begin{eqnarray}
  N_3
     &=&
        \sum_{\lambda_3=0}^{k_3 - 1}
        \sum_{M=2}^{\infty}
                \sum_{l=1}^{\infty}
                \frac{(M-1)M}{2}
                z^l e^{-l (M \eta_1 + \lambda_3 \eta_3) }
      \nonumber  \\
%    &=&
%       \frac{ 1 }{ 2 }
%       \sum_{l=1}^{\infty}
%       z^l
%       \left[
%           \frac{ e^{-l \eta_1} ( 1 + e^{-l \eta_1}) }
%           { ( 1 - e^{-l \eta_1} )^2  (1 - e^{-l \eta_3} ) }
%           -
%           \frac{ e^{-l \eta_1}  }
%           { ( 1 - e^{-l \eta_1} )  (1 - e^{-l \eta_3} ) }
%       \right]
%    \nonumber  \\
         &=&
        \sum_{l=1}^{\infty}
        z^l
        \frac{ e^{-2 l \eta_1} }
        { ( 1 - e^{-l \eta_1} )^2  (1 - e^{-l \eta_3} ) }
     \nonumber  \\
     &=&
            \sum_{l=1}^{\infty}
            \frac{ z^l e^{- l (\eta_1 - \eta_3 / 2) }}
        { l^3 \eta_1^{~2} \eta_3 }
        -
            \sum_{l=1}^{\infty}
            \frac{ z^l e^{- l (\eta_1 - \eta_3 / 2) }}
        { 12 l \eta_3 }
         + \cdots
     \nonumber  \\
     &=&
        \frac{  g_3( z e^{- (\eta_1 - \eta_3 / 2) }  ) }{ \eta_1^{~2} %%@
\eta_3 }
        -
        \frac{  g_1( z e^{- (\eta_1 - \eta_3 / 2) }  ) }{ 12 \eta_3 }
         + \cdots.
                    \label{300}
\end{eqnarray}
%
%%%%%%%%%%%%%%%%%%%%%%%%%%%%%%%%%%%%%%%%%%%%%%5
%
%     Oblate Shape Potential
%
%%%%%%%%%%%%%%%%%%%%%%%%%%%%%%%%%%%%%%%%%%%%%%%

%%%%%%%%%%%%%%%%%%%%%%%
\subsubsection{Oblate shape potential}

Next we discuss the case of anisotropy
 corresponding to an oblate shape potential
$\omega_1 > \omega_2 = \omega_3$
( $k_1 = 1 < k_2 = k_3$).
%For a strong anisotropy $k_2 = k_3 >> 1$,
%one can observe two-dimensional condensation.
In this case, $\omega = \omega_1$ and
\begin{equation}
    \bar{E}_n =\omega M + \omega_2 (\lambda_2 + \lambda_3).
      \label{Oc3f}
\end{equation}

 The number of particles excited
 in these dimensions are given respectively by
\begin{eqnarray}
  N_0 &=& \frac{z}{1-z},
    \\        \label{Of90}
  N_1 &=& \sum_{n_2=1}^{\infty} \frac{2 z}{e^{ n_2 \eta_2 } - z},
            \label{Of91}  \\
  N_2 &=& \sum_{n_2=1,n_3=1}^{\infty}
  \frac{z}{e^{ n_2 \eta_2 + n_3 \eta_3 } - z}
    \nonumber  \\
        &=& \sum_{M=2}^{\infty}
  \frac{(M - 1) z}{e^{ M \eta_2 } - z},
              \label{Of92} \\
  N_3 &=& \sum_{n_1=1,n_2=0,n_3=0}^{\infty}
  \frac{z}{ e^{ n_1 \eta_1 + n_2 \eta_2 + n_3 \eta_3 } - z}
    \nonumber  \\
        &=& \sum_{n_1=1}^{\infty}  \sum_{M=0}^{\infty}
  \frac{(M+1) z}{ e^{ n_1 \eta_1 + M \eta_2 } - z},
                    \label{Of93}
\end{eqnarray}
where the factor $2$ in Eq. (\ref{Of91}) is due to the symmetry between
the second and third axis.

%%%%%%%%%%%%%%%%%%%%%%%
%
% N1
%
%%%%%%%%%%%%%%%%%%%%%%%
Compared to Eq. (\ref{f10}),
we obtain
\begin{eqnarray}
  N_1 &=&
    \frac{ 2 g_1(z e^{ - \eta_2 / 2 } ) }{ \eta_2 }
      + \cdots
\end{eqnarray}
in the present case.
%%%%%%%%%%%%%%%%%%%%%%%
%
% N2
%
%%%%%%%%%%%%%%%%%%%%%%%

The number of particles excited two-dimensionally on $x_2-x_3$ plane
can be written as
\begin{eqnarray}
    N_2
     &=&
     \sum_{l=1}^{\infty}
         \sum_{M=2}^{\infty}
         (M-1) z^l e^{-l M \eta_2}
     \nonumber  \\
     &=&
         \sum_{l=1}^{\infty}
        \frac{ z^l e^{-2 l \eta_2}}
        { ( 1 - e^{-l \eta_2} )^2 }
     \nonumber  \\
     &=&
     \frac{ g_2( z  e^{- \eta_2}) }{ \eta_2^{~2} }
         + \cdots.
        \label{Of15}
\end{eqnarray}

%%%%%%%%%%%%%%%%%%%%%%%
%
% N3
%
%%%%%%%%%%%%%%%%%%%%%%%

For three dimensional excitations, Eq. (\ref{Of93}) gives
\begin{eqnarray}
  N_3
     &=&
        \sum_{l=1}^{\infty}
        z^l
        \frac{ e^{- l \eta_1} }
        { ( 1 - e^{-l \eta_1} )  (1 - e^{-l \eta_2} )^2  }
     \nonumber  \\
     &=&
    \frac{  g_3( z  e^{- ( \eta_1/2 - \eta_2) }  ) }
         { \eta_1 \eta_2^{~2} }
        -
    \frac{  g_1( z  e^{- ( \eta_1/2 - \eta_2) }  ) }{ 24 }
                ( \frac{ \eta_1^{~2} + 2 \eta_2^{~2} }
               { \eta_1 \eta_2^{~2} }
                 )
        + \cdots.
        \label{O300}
\end{eqnarray}
%%%%%%%%%%%%%%%%%%%%%%%%%%%%%%%%%%%%%%%%%%%%%%%%%%%
%
%    Maximally Anisotropic Potential
%
%%%%%%%%%%%%%%%%%%%%%%%%%%%%%%%%%%%%%%%%%%%%%%%%%%%

\subsubsection{Maximally anisotropic potential}

For anisotropies
 $\omega_1 > \omega_2 > \omega_3$ with
 $k_1 =1 << k_2 << k_3$,
 $\omega = \omega_1$ and
\begin{equation}
    \bar{E}_n =
    \omega M + \omega_2 \lambda_2 + \omega_3 \lambda_3 .
      \label{c3f61}
\end{equation}

 The number of excited modes
  in the corresponding dimensions can be defined by
\begin{eqnarray}
  N_1 &=& \sum_{n_3=1}^{\infty} \frac{z}{e^{ n_3 \eta_3 } - z},
    \\          \label{f910}
  N_2 &=& \sum_{n_2=1, n_3=0}^{\infty}
  \frac{z}{e^{ n_2 \eta_2 + n_3 \eta_3 } - z},
           \label{f920}
\\
  N_3 &=& \sum_{n_1=1,n_2=n_3=0}^{\infty}
  \frac{z}{ e^{ n_1 \eta_1 + n_2 \eta_2 + n_3 \eta_3 } - z}
    \nonumber  \\
  &=&
       \sum_{\lambda_2=0}^{k_2 - 1}
       \sum_{\lambda_3=0}^{k_3 - 1}
       \sum_{M=1}^{\infty}
  \frac{M(M+1)}{2}
  \frac{z}{ e^{ M  \eta_1 + \lambda_2 \eta_2 + \lambda_3 \eta_3 } - z}.
            \label{f930}
\end{eqnarray}
%

%%%%%%%%%%%%%%%%%%%%%%%
%
% N2
%
%%%%%%%%%%%%%%%%%%%%%%%

For the two dimensional case, following Eq. (\ref{f15}),
\begin{eqnarray}
    N_2
     &=&
         \sum_{\lambda_3=0}^{\kappa - 1}  \sum_{M=1}^{\infty}
         \sum_{l=1}^{\infty} M z^l e^{-l (M \eta_2 + \lambda_3 \eta_3) }
     \nonumber  \\
     &=&
         \sum_{l=1}^{\infty}
        \frac{ z^l e^{-l \eta_2}}
        { ( 1 - e^{-l \eta_2} ) ( 1 - e^{-l \eta_3} ) }
     \nonumber  \\
     &=&
         \sum_{l=1}^{\infty}
         \frac{ z^l e^{- \eta_2 / 2 }}
        { l^2 \eta_2 \eta_3 }
    -
            ( \kappa + \frac{1}{\kappa} )
               \sum_{l=1}^{\infty}
               \frac{ z^l e^{   - \eta_2 / 2 }}
                { 24 }
         + \cdots
     \nonumber  \\
     &=&
        \frac{ g_2( z e^{- \eta_2 / 2 }  )}{ \eta_2 \eta_3 }
        -
        \frac{ \kappa g_0( z e^{- \eta_2 / 2 }  )}{ 24 }
         + \cdots,
        \label{f150}
\end{eqnarray}
where $\kappa \equiv k_3/k_2$.

%%%%%%%%%%%%%%%%%%%%%%%
%
% N3
%
%%%%%%%%%%%%%%%%%%%%%%%

For three dimensional excitations, Eq. (\ref{f930}) gives
\begin{eqnarray}
  N_3
 &=&
%       \frac{ 1 }{ 2 }
%       \sum_{l=1}^{\infty}
%       z^l
%       \left[
%           \frac{ e^{-l \eta_1} ( 1 + e^{-l \eta_1}) }
%           { ( 1 - e^{-l \eta_1} )^3 }
%           +
%           \frac{ e^{-l \eta_1} }
%           { ( 1 - e^{-l \eta_1} )^2 }
%       \right]
%       \left(
%           \frac{ 1 - e^{-l \eta_1} }
%                { 1 - e^{-l \eta_2} }
%           \right)
%       \left(
%           \frac{ 1 - e^{-l \eta_1} }
%                { 1 - e^{-l \eta_3} }
%           \right)
%    \nonumber  \\
%        &=&
        \sum_{l=1}^{\infty}
        z^l
        \frac{ e^{- l \eta_1} }
    {  ( 1 - e^{-l \eta_1} ) ( 1 - e^{-l \eta_2} ) (1 - e^{-l \eta_3} ) }
     \nonumber  \\
     &=&
            \sum_{l=1}^{\infty}
            \frac{ z^l e^{- l ( \eta_1 - \eta_2 - \eta_3) / 2} }
        { l^3 \eta_1 \eta_2 \eta_3 }
        -
            \sum_{l=1}^{\infty}
            \frac{ z^l  e^{- l ( \eta_1 - \eta_2 - \eta_3) / 2 } }
        { 24 l }
                ( \frac{ \eta_1^{~2} + \eta_2^{~2} + \eta_3^{~2} }
               { \eta_1 \eta_2 \eta_3 }
                 )
         + \cdots
     \nonumber  \\
     &=&
         \frac{ g_3( z  e^{- ( \eta_1 - \eta_2 - \eta_3) / 2 }  ) }
         { \eta_1 \eta_2 \eta_3 }
        -
     \frac{ g_1( z  e^{- ( \eta_1 - \eta_2 - \eta_3) / 2}  ) }{ 24 }
                ( \frac{ \eta_1^{~2} + \eta_2^{~2} + \eta_3^{~2} }
               { \eta_1 \eta_2 \eta_3 }
                  )
         + \cdots.
                    \label{3002}
\end{eqnarray}

\section{Finite Size Effects and Dimensional Crossover Behavior}

\subsection{Bulk behavior}

%%%%%%%%%%%%%%%%%%%%%%%
%
% T3d
%
%%%%%%%%%%%%%%%%%%%%%%%

The bulk three-dimensional condensation temperature is defined
in the thermodynamic limit
$\eta_i \rightarrow 0 ~(i=1,2,3)$ and
$N \rightarrow \infty$ while $\eta_1 / \eta_3$ and $\eta_2 / \eta_3$ fixed.
The dominant term is given by the first term in $N_3$ and
the critical temperature satisfies
\begin{eqnarray}
    N =
    \frac{ T_{3D}^{~3} }{ \omega_1 \omega_2 \omega_3 } \zeta(3).
     \label{T3d}
\end{eqnarray}
Therefore
\begin{eqnarray}
   T_{3D} = \left( \frac{N \omega_1 \omega_2 \omega_3 }{ \zeta(3) }
            \right)^{1/3}.
    \label{T3d2}
\end{eqnarray}
%%%%%%%%%%%%%%%%%%%%%%%
%
% T2d
%
%%%%%%%%%%%%%%%%%%%%%%%
Two-dimensional limit is given by $\eta_2,\eta_3 \rightarrow 0$,
$N \rightarrow \infty$, but $\eta_1 >> 1$.
In such a case, the dominant term in particle number is $N_2$ and
the critical temperature $T_{2D}$ is defined by
\begin{eqnarray}
    N = \frac{T_{2D}^{~2}}{ \omega_2 \omega_3 }
    g_2(e^{ - (\omega_2 + \omega_3) / 2 T_{2D}} )
    =
    \frac{T_{2D}^{~2}}{ \omega_2 \omega_3 }
    \zeta(2) + \cdots.
    \label{T2d}
\end{eqnarray}
Thus we have
\begin{eqnarray}
   T_{2D} = \left( \frac{N \omega_2 \omega_3}{ \zeta(2) }
            \right)^{1/2}.
    \label{T2d2}
\end{eqnarray}
%
%%%%%%%%%%%%%%%%%%%%%%%
%
% T1d
%
%%%%%%%%%%%%%%%%%%%%%%%
For one-dimensional limit $\eta_3 \rightarrow 0$,
$N \rightarrow \infty$, but $\eta_1,\eta_2 >> 1$,
the dominant term in particle number is $N_1$.
The condensation temperature is defined by
\begin{eqnarray}
    N = \frac{T_{1D}}{ \omega_3 }
    g_1(e^{ - \omega_3 / 2 T_{1D}} ).
    \label{T1d}
\end{eqnarray}
To leading order in $\eta_3^{~-1}$, Eq. (\ref{f10})
can be approximated as $ N_1 \sim \frac{T}{ \omega_3 }
 \log \frac{2 T}{ \omega_3 }  $.
Thus the one-dimensional condensation temperature
$T_{1D}$ is defined by
\begin{eqnarray}
    N = \frac{T_{1D}}{ \omega_3 }
      \log \frac{2 T_{1D}}{ \omega_3 },
    \label{f101a}
\end{eqnarray}
which gives
\begin{eqnarray}
   T_{1D} = \frac{N \omega_3}{ \log(2N)}
    \label{f101b}
\end{eqnarray}
 for large $N$.
Note that in the thermodynamic limit $N \rightarrow \infty$
while $N \omega_3$ fixed, $T_{1D}$ vanishes.
%In Figure 8, we plot $T_{1D}^{(0)}$ and $T_{1D}$ for various $N$.

%%%%%%%%%%%%%%%%%%%%%%%%%%%%%%%%%%%%%%%%%%%%%%%%%%%%%%%%%%%%%%%%
In Figure 1, we plot $T_{1D}$, $T_{2D}$, and $T_{3D}$ as a
function of the anisotropy parameter $k_3$. In an isotropic and
weakly isotropic case ($k_3 \sim 1$), $T_{3D} < T_{1D},T_{2D}$
and the condensation is directly into the ground state. As
anisotropy is increased ($k_3 >> 10^3$), $T_{1D}, T_{2D} < T_{3D}$
is achieved. This is the regime where various multistep behaviors
can take place.

%For higher anisotropy ($k_3 \sim 10^4$ and $k_2 \sim 10^2$),
%$T_{1D} < T_{2D} < T_{3D}$ is achieved and three-step behavior
%can be observed.
%%%%%%%%%%%%%%%%%%%%%%%%%%%%%%%%%%%%%%%%%%%%%%%%%%%%%%%%%%%%%%%%

%
%\begin{figure}[h]
%  \begin{center}
%\epsfxsize=.6\textwidth \epsfbox{LFig1.eps}
%  \end{center}
%\caption{
%The condensation temperatures $T_{1D}$ (solid curve),
% $T_{2D}$ (dashed curve), and
% $T_{3D}$ (dot-dashed curve) are shown
%as a function of $k_3$. $N=10^4$,
%$\omega_1=0.5$, $k_2$ is also varied as $k_2^{~2}=k_3$.
%The logarithmic scale is used for both axis.
%}
%\label{fig1}
%\end{figure}
%
In terms of bulk condensation temperatures we obtained in Section
3.1 as
 $T_{1D} = N \omega_3 / \log(2 N)  $,
 $T_{2D} = ( N \omega_2 \omega_3 / 2 \zeta(2) )^{1/2} $, and
 $T_{3D} = ( N \omega_1 \omega_2 \omega_3 / \zeta(3) )^{1/3} $,
% in Eqs. (\ref{f101b}), (\ref{T2d2}), and (\ref{T3d2})
the conditions
 (A)  $T_{1D} << T_{2D}$,
 (B)  $T_{2D} << T_{3D}$, and
 (C)  $T_{1D} << T_{3D}$
 give constraints for
 $k_3$ and $\kappa$ as
\begin{eqnarray}
   (A)& \kappa  &>>
         \frac{ N \zeta(2) }{ (\log(2N))^2 },
%   \mbox{ ~( $ T_{1D}^{(0)} < T_{2D}^{(0)} $) },
    \nonumber  \\
   (B)& k_3^{~2} / \kappa &>>
              \frac{ N \zeta(3)^2 }{ \zeta(2)^3 },
%   \mbox{ ~~~~~~~~( $ T_{2D}^{(0)} < T_{3D}^{(0)} $) },
        \nonumber  \\
   (C)&  k_3 \kappa &>>
              \frac{ N^2 }{ (\log(2 N))^3 }.
%   \mbox{ ~~~~~~~~( $ T_{2D}^{(0)} < T_{3D}^{(0)} $) }
        \label{f18}
\end{eqnarray}
Since (B) $T_{2D} << T_{3D}$ implies $T_{3D} << \omega_1$,
three-step BEC never occurs in harmonic traps.
 In Figure 2, different condensation behaviors corresponding to
various anisotropy parameters
 $\omega_1 / \omega_2$ and $\omega_2 / \omega_3$ are shown.
The vertical axis corresponds to the prolate-shape potential
studied in Section 3.2.1 (this case was studied in \cite{DruKet97}).
In such a potential, two-step BEC can be seen.
The horizontal axis corresponds to the oblate-shape potential
discussed in Section 3.2.2, where we show that
there is no multistep condensation in this case.
The more general class of anisotropic case will
be discussed in Section 3.2.3.
The combined effect of dynamical dimensional
reduction and two-step BEC in such a potential can appear in three steps.
%%%%%%%%%%%%%%%%%%%%%%%
%
%   FSE & Crossover
%
%%%%%%%%%%%%%%%%%%%%%%%

\subsection{Dimensional crossover and condensation}

For a highly anisotropic trap,
the three dimensional crossover temperature
$T_{3D}^{*}$ should be reached when the correlation length is in the order
of the size of the ground state wave function in the most confining %%@
direction.
Spreading of the wave function can be characterized by
$L_i \equiv \sqrt{ \hbar / m \omega_i }$ (for $i=1,2,3$)
\cite{BayPet96,CYY96}.
Hence the above condition is equivalent to $\xi(T_{3D}^{*}) \sim
\lambda_{\theta dB} / \sqrt{t_3} \sim
L_1$,
where $\lambda_{\theta dB} \equiv h / \sqrt{2\pi m k T}$
is the thermal de Broglie wavelength, and
$t_3 \equiv  | T_{3D}^{*} - T_{3D} | / T_{3D} $.
This will give us the crude estimate of $T_{3D}^{*}$ as
\begin{eqnarray}
  | T_{3D}^{*} - T_{3D} |
  \sim \left( \frac{ k_3 \zeta(3)}{ N }  \right)^{1/3} T_{3D}.
    \label{t3d}
\end{eqnarray}

\subsubsection{Two-step condensation}

%%%%%%%%%%%%%%%%%%%%%%%
%
%   FSE & Crossover T3*
%
%%%%%%%%%%%%%%%%%%%%%%%

For a prolate shape potential discussed in Section 2.1.1,
we expand the whole particle spectrum
with respect to $\eta_1$ and $\eta_3$  and obtain
\begin{eqnarray}
  N  &=&
        N_0 + N_1 + N_2 + N_3
  \nonumber  \\
     &=&
     g_0( z ) +
     \frac{ g_1( z e^{- \eta_3 / 2}  ) }{ \eta_3 }
     +
     \frac{ 2 g_2( z e^{- \eta_1 / 2}  ) }{ \eta_1 \eta_3 }
     +
     \frac{ g_3( z e^{- \eta_1 }  ) }{ \eta_1^{~2} \eta_3 }
         + \cdots
  \nonumber  \\
     &=&
     g_0( z ) +
     \frac{ g_1( z  ) }{ \eta_3 }
     +
     \frac{ 2 g_2( z ) }{ \eta_1 \eta_3 }
         + \frac{ g_3( z ) }{ \eta_1^{~2} \eta_3 }
  \nonumber  \\
         &-& \frac{ g_0( z ) }{ \eta_3 }
         \frac{ \eta_3 }{ 2 }
       - \frac{ 2 g_1( z ) }{ \eta_1 \eta_3 }
         \frac{ \eta_1 }{ 2 }
       - \frac{ g_2( z ) }{ \eta_1^{~2} \eta_3 }
         \eta_1
  \nonumber  \\
       &+& \frac{ g_0( z ) }{ \eta_1 \eta_3  }
         \left( \frac{ \eta_1 }{ 2 } \right)^2
       + \frac{ g_1( z ) }{ \eta_1^{~2} \eta_3 }
         \frac{ \eta_1 ^2 }{ 2 }
       - \frac{ g_0( z ) }{ \eta_1^{~2} \eta_3  }
         \frac{ \eta_1^{~3} }{ 6 }
       + \cdots.
     \label{Ntotal}
\end{eqnarray}
This expression can be simplified to give
\begin{eqnarray}
  N \eta_3 =
    \frac{ g_3( z )  }{ \eta_1^{~2}  }
    +
    \frac{ g_2( z )  }{ \eta_1 }
    +
    \frac{ g_1( z )  }{ 2 }
    + \cdots.
     \label{Ntotal2}
\end{eqnarray}
Writing $z=e^{-\phi}$ and expanding Eq. ({\ref{Ntotal2})
with respect to $\phi$ give
\begin{eqnarray}
  N \eta_3 =
     \frac{ \zeta(3)  }{ \eta_1^{~2}  }
+    \frac{ \zeta(2)  }{ \eta_1 }
-    \frac{ \zeta(2) \phi  }{  \eta_1^{~2}  }
+ \cdots
     \label{Ntotal3},
\end{eqnarray}
where we used an asymptotic expansion of the Bose-Einstein function
 $ g_3( e^{- \alpha } ) \sim
 \zeta(3) - \zeta(2) \alpha + 1/2 ( 3/2 - \log \alpha)
 \alpha^2 + \cdots $ and
 $ g_2( e^{- \alpha } ) \sim \zeta(2) + ( \log \alpha - 1) \alpha + \cdots$
 for small $\alpha $
 \cite{Pathria}.

Correlation length $\xi$ of an ideal Bose gas is given by
$\xi = \lambda_{\theta dB} / 2 \sqrt{\pi \phi}$ \cite{GunBuc68}.
In terms of scaling parameters
$x_i(T) \equiv \phi(T) / \eta_i (i = 1,2,3)$,
the above argument implies that $T_{3D}^{*}$ is
achieved when $x_1(T_{3D}^{*}) \equiv c_1$, where $c_1$ is some constant
in the order of unity.
Inserting this into Eq. (\ref{Ntotal3}) gives
\begin{eqnarray}
  N =
   \frac{ \zeta(3)  }{ \eta_1^{~2} \eta_3 }
   +
   \frac{ \zeta(2) (1 - c_1) }{ \eta_1 \eta_3 }
   + \cdots
   ~~{\mbox at} ~~T = T_{3D}^{*}.
     \label{Ntotal4}
\end{eqnarray}
Thus we have
\begin{eqnarray}
 \frac{ T_{3D}^{*} }{ T_{3D} } = 1 +
 \frac{ c_1 - 1 }{ 3 } \frac{ \zeta(2) }{ \zeta(3)^{~2/3} }
 \left( \frac{ k_3 }{ N }  \right)^{1/3}.
     \label{Ntotal5}
\end{eqnarray}
This result gives the same correction term proportional to
$\left( k_3 / N   \right)^{1/3}$
as in Eq. (\ref{t3d})
obtained by heuristic arguments.

%%%%%%%%%%%%%%%%%%%%%%%
%
%   FSE & Crossover T1D*
%
%%%%%%%%%%%%%%%%%%%%%%%

The one-dimensional condensation temperature $T_{1D}$ is defined in
Eq. (\ref{T1d}) in the limit of small $\eta_3$ and the vanishing
reduced chemical potential $\phi$;
finite size effects on $T_{1D}$ originate in
finiteness of both $\eta_3$ and $\phi$.
At one-dimensional crossover temperature $T_{1D}^{*}$,
correlation length reaches the size of the ground state wave function
in the least confining direction, namely,
along the third axis in the present case.
Thus
$\xi(T_{1D}^{*}) \sim L_3$ or equivalently
$x_3(T_{1D}^{*}) \equiv c_3 = O(1)$.
Then from the second line in Eq. (\ref{Ntotal}),
we obtain
\begin{eqnarray}
  N  &=&
     \frac{ g_1( e^{- (1 + 2 c_3) \eta_3 / 2}  ) }{ \eta_3 }
     +
     \frac{ 2 g_2( e^{- (k_3 + 2 c_3) \eta_3 / 2}  ) }{ \eta_1 \eta_3 }
     +
     \frac{ g_3( e^{- (k_3 + c_3) \eta_3}  ) }{ \eta_1^{~2} \eta_3 }
         + \cdots.
%
% \nonumber  \\
% &=&
%
%    -
%    \frac{ \log( 1 - e^{- (1 + 2 c_3) \eta_3 / 2}  ) }{ \eta_3 }
%    +
%    \frac{ 2 e^{- \eta_1 / 2} }{ \eta_1 \eta_3 }
%    +
%    \frac{ e^{- \eta_1 } }{ \eta_1^{~2} \eta_3 }
%         + \cdots.
%
     \label{Ntotal1d}
\end{eqnarray}
In the limit $\eta_3 \rightarrow 0$,
%For strong anisotropy $k_3 >> 1$,
%large $\eta_1$,
only the first term dominates and we obtain
the crossover temperature as
\begin{eqnarray}
    N &=& - \frac{ T_{1D}^{*} }{ \omega_3 }
    \log( 1 - e^{- (1 + 2 c_3) \omega_3 / 2 T_{1D}^{*} }  )
    \nonumber  \\
 &=&   \frac{ T_{1D}^{*} }{ \omega_3 }
      \log \frac{2 T_{1D}^{*} }{ (1 + 2 c_3) \omega_3 }.
    \label{T1dcvr1}
\end{eqnarray}
For large $N$,
this gives
\begin{eqnarray}
   T_{1D}^{*} = \frac{N \omega_3}{ \log[2N/(1 + 2 c_3)]}.
    \label{T1dcvr2}
\end{eqnarray}
 Note that $ T_{1D}^{*} > T_{1D} $ holds.

The conditions (C) in Eq. (\ref{f18}) and
$\omega_1 < T_{1D} $ are satisfied if
\begin{eqnarray}
\frac{ N }{ (\log(2 N))^{3/2} }
 <  k_3 < \frac{ N }{ (\log(2 N)) }.
    \label{2step}
\end{eqnarray}
In such a case, two-step condensation leading to the condensation
into the ground state can be seen whereas the system is
effectively still three-dimensional. In Figure 3-6, the
condensation fractions
 $ N_i / N (i=0,1,2,3) $ as a function of
 temperature are plotted.
 At high temperature, three-dimensionally excited states dominate,
 as expected from the density of states which grows as $M^2$,
 where $M$ is the number of degeneracy of an isotropic harmonic oscillator
 appeared in Eq. (\ref{f2}).
% If we lower the temperature from above, at $T_{3D}$,
% three-dimensionally excited states start condensing into two-dimensionally
% excited states.
% However, at this stage, one-dimensionally excited states and
% ground state contributions are small.
% Further lowering temperature to $T_{2D}$, two-dimensionally excited
% states start condensing into one-dimensionally excited states.
% At $T_{1D}$, one-dimensionally excited states saturate,
% and the condensation into ground state sets in.

In the isotropic case (Figure 3),
condensation is only into the ground state.
Due to the finite size effects, condensation already starts
before the critical temperature is reached.
In a strongly anisotropic case, as in Figure 4,
two-step condensation can be seen.
$T_{3D}^{*}$ determines the onset of condensation
into one-dimensionally excited states.
At $T_{3D}^{*}$ the ground state fraction is negligiblly small.
Condensation into the ground state will not start until
$T_{1D}^{*}$ is reached.

In the multistep process peculiar to the highly anisotropic system,
 when the correlation length reaches the size of
the system, the dynamics shows the crossover to the
low-dimensional one before the actual phase transition occurs. In
this sense, the critical temperature is never observed in such a
process and the directly relevant quantity to the observation is
the crossover temperature, the temperature at which the finite
size correction sets in. For practical purposes, this is often
replaced by including the finite size correction as the term
proportional to the power of $1/N$ whereas the chemical potential
is set to the ground state energy \cite{KetDru96}. Strictly
speaking, however, since the chemical potential never reaches the
ground state energy in the finite system, the meaning of this
correction has some ambiguity. The difference between the
crossover temperature and the finite size-corrected critical
temperature in the present axially symmetric trap case is given by
\begin{eqnarray}
 \frac{ \Delta T }{ T_{3D} } =
 \frac{ c_1 }{ 3 } \frac{ \zeta(2) }{ \zeta(3)^{~2/3} }
 \left( \frac{ k_3 }{ N }  \right)^{1/3}.
     \label{DIFF}
\end{eqnarray}
While this is fairly small for an isotropic or weakly anisotropic case:
$\Delta T / T_{3D} \sim 0.024    $ for $k_3=1$,
it is no longer so for a strongly anisotropic case:
$\Delta T / T_{3D} \sim 0.24    $ for $k_3=10^3$ as used in Figure 4,
where $c_1=1, N=10^4$ for both cases.
The ordinary finite size correction significantly underestimates the
results in the latter case.\footnote{Since the boundary condition in the harmonic potential
corresponds to the one in the Neumann boundary condition,
the surface correction increases
the density of states and hence decreases the critical temperature. }
For these reasons, we focus our discussions on crossover temperatures
in the present work. 
We should also note that there is a slight amount of ambiguity in the choice of $c_i  (i=1,2,3)$.
In general, the correlation length $\xi$ is a complicated function of the temperature away from the critical value and 
the reliable choice is obtained by 
the numerical fitting. We will simply put $c_i  (i=1,2,3) = 1$ 
for our comparison with numerical data for brevity.

%%%%%%%%%%%%%%%%%%%%%%%
%
%   FSE & Crossover: Two-dimensional
%
%%%%%%%%%%%%%%%%%%%%%%%

\subsubsection{Two-dimensional condensation}

For an oblate shape potential discussed in Section 2.1.2,
assuming $\eta_1 >> 1$ and we obtain
\begin{eqnarray}
  N  &=&
        N_0 + N_1 + N_2 + N_3
 \nonumber  \\
  &=&
         g_0( z )
         + \frac{2  g_1( z e^{- \eta_2 / 2}  ) }{ \eta_2 }
%    - \frac{\kappa}{24} g_0( z  e^{- \eta_2 / 2} )
         + \frac{ g_2( z e^{- \eta_2}  ) }{ \eta_2^{~2} }
%    + \frac{ g_3( z e^{- \eta_1 / 2 + \eta_2}  ) }{ \eta_1 \eta_2^{~2} }
%\nonumber  \\
%  &-& \frac{ g_1( z e^{- \eta_1 / 2 + \eta_2}  ) }{ 24 }
%         ( \frac{ \eta_1^{~2} + 2 \eta_2^{~2} }
%              { \eta_1 \eta_2^{~2} }
%         )
         + \cdots.
% \nonumber  \\
%  &=&
%           \frac{ g_3( z ) }{ \eta_1 \eta_2^{~2} }
%         + \frac{ g_2( z ) }{ \eta_2^{~2}}
%    - \frac{ g_2( z ) }{ \eta_1 \eta_2^{~2}}
%           \left( \frac{ \eta_1 }{ 2 } - \eta_2
%           \right)
%         + \cdots.
         \label{Ntotal2dcond}
\end{eqnarray}
%
%Writing $z=e^{-\phi}$ and expanding Eq. ({\ref{Ntotal2dcond})
%with respect to $\phi$ gives
%%
%\begin{eqnarray}
%%
%  N &=&
%    \frac{ \zeta(3)  }{ \eta_1 \eta_2^{~2} }
%+   \frac{ \zeta(2)  }{ 2 }
%\left(  \frac{ 2 }{ \eta_1 \eta_2 }
% + \frac{ 1 }{ \eta_2^{~2}}
%\right)
% - \frac{ \zeta(2) \phi  }{  \eta_1 \eta_2^{~2}} + \cdots.
%     \label{Ntotal2d2}
%
%\end{eqnarray}
%
%As discussed in Section 3.2.1,
%at $T_{3D}^{*}$,
%$x_1(T_{3D}^{*}) \equiv c_1 = O(1)$ holds.
%
%Thus we have
%
%\begin{eqnarray}
%
% \frac{ T_{3D}^{*} }{ T_{3D} } = 1 +
% \frac{ c_1 - 1/2 }{ 3 } \frac{ \zeta(2) }{ \zeta(3)^{~2/3} }
% \left( \frac{ k_3^{~2} }{ N }  \right)^{1/3}.
%     \label{Ntotal2d5}
%
%\end{eqnarray}
%

%%%%%%%%%%%%%%%%%%%%%%%
%
%   FSE & Crossover T2D*
%
%%%%%%%%%%%%%%%%%%%%%%%

Two-dimensional crossover temperature $T_{2D}^{*}$ can be defined
when the correlation length reaches the size of the ground state wave %%@
function
along the second axis.
Thus we have $\xi(T_{2D}^{*}) \sim L_2$ or equivalently
$x_2(T_{2D}^{*}) \equiv c_2 = O(1)$.
Then from Eq. (\ref{Ntotal2dcond}) we obtain
\begin{eqnarray}
  N  &=&
     \frac{ g_2( e^{- ( 1 + c_2 ) \eta_2 } ) }{ \eta_2^{~2} }
        + \frac{ 2 g_1( e^{- ( 1 + 2 c_2 ) \eta_2 / 2 } ) }{ \eta_2 }
%        + \frac{ e^{- \eta_1 / 2 } }{ \eta_1 \eta_2^{~2} }
%   +
%    \frac{ \log( 1 - e^{- \eta_1 / 2}  ) \eta_1 }{ 24 \eta_2^{~2}}
     + \cdots
     \label{Ntotal2d22}
\end{eqnarray}
at $T = T_{2D}^{*}$.
%For $k_3 >> 1$,
%For large $\eta_1$,
%only the first two terms dominate
Expanding in terms of $\eta_2$ and we obtain $T_{2D}^{*}$ as
\begin{eqnarray}
    N &=& \frac{ \zeta(2) }{ \eta_2^{~2}}
    - \frac{ 1 + c_2 }{ \eta_2 }
    + \frac{ (1 + c_2) \log [\eta_2 (1 + c_2)] }{ \eta_2 }
    - \frac{ 2 \log (\eta_2 c_2) }{ \eta_2 }
%       ~~{\mbox at} ~~T = T_{2D}^{*}
    \nonumber  \\
 &=&
    \frac{ T_{2D}^{*2} \zeta(2) }{ \omega_2^{~2}}
 -  \frac{ T_{2D}^{*} }{ \omega_2 }
    \left[
     (1 + c_2) \right.
   \nonumber  \\
  &+&
    \left.
    (1 + c_2) \log \left( \frac{ T_{2D}^{*} }{\omega_2 (1 + c_2)} \right)
    + 2 \log \left( \frac{ T_{2D}^{*} }{\omega_2 c_2} \right)
%%%%%%%%%%%%%%%%%%%%%%%%%%%%%%%%%%%%%%%%%%%%%%%%%%%%%%%%%%%%%%%%%%%%%%%%
%
    \right] + \cdots
    \label{T2d2cvr}
\end{eqnarray}
for $\eta_2, \eta_3 << 1$.
For large $N$ and $c_2=1$, this gives
\begin{eqnarray}
 \frac{ T_{2D}^{*} }{ T_{2D} }
 &=&
  1 +
 \left( \frac{ \kappa }{ N \zeta(2) }  \right)^{1/2}
\log
     \left( \frac{ N }{ \kappa \zeta(2) } \right).
     \label{Ntotal2d52}
\end{eqnarray}

As explained in Section 3.1, condensation into $N_2$ does not
occur in harmonic traps. For an oblate shape potential, the system
dynamics freezes out along the first axis at $T = \omega_1$.
Therefore the dynamics of the system at $T < \omega_1$ is
two-dimensional. Ordinary two-dimensional BEC can still be
observed as long as $T_{2D} < \omega_1$. This condition requires
$k_3 > (N / \zeta(2))^{1/2}$. In Figure 5, two-dimensional BEC in
this parameter regime is shown.

%However, if we start from $EIRD=2$, two-step behavior
%, $N_2$ to $N_1$, $N_1$ to $N_0$ can be realized.
%When $T_{2D} < \omega_1 < T$,
%dimensional reduction occurs at $T \sim \omega_1$ followed
%by two-step condensation at $T_{2D}$ and $T_{1D}$.
%In Figure 5, two-dimensional condensation is plotted.

%%%%%%%%%%%%%%%%%%%%%%%
%
%   FSE & Crossover: Three-Step
%
%%%%%%%%%%%%%%%%%%%%%%%

\subsubsection{Three-step dimensional reduction}

For a general class of anisotropic potential discussed in Section
2.1.3, three-step process can be observed. We expand each $N_i$
with respect to $\eta_1,\eta_2,\eta_3$ and obtain
\begin{eqnarray}
  N  &=&
        N_0 + N_1 + N_2 + N_3
 \nonumber  \\
  &=&
         g_0( z )
         + \frac{ g_1( z e^{- \eta_3 / 2}  ) }{ \eta_3 }
         + \frac{ g_2( z e^{- \eta_2 / 2}  ) }{ \eta_2 \eta_3 }
     - \frac{\kappa}{24} g_0( z  e^{- \eta_2 / 2} )
         + \frac{ g_3( z e^{- (\eta_1 - \eta_2 - \eta_3) / 2}  ) }
{ \eta_1 \eta_2 \eta_3 }
\nonumber  \\
  &-& \frac{ g_1( z e^{- (\eta_1 - \eta_2 - \eta_3) / 2}  ) }{ 24 }
         ( \frac{ \eta_1^{~2} + \eta_2^{~2} + \eta_3^{~2} }
               { \eta_1 \eta_2 \eta_3 }
         )
         + \cdots
 \nonumber  \\
  &=&
           \frac{ g_3( z ) }{ \eta_1 \eta_2 \eta_3 }
         + \frac{ g_2( z ) }{ \eta_2 \eta_3 }
     - \frac{ g_2( z ) }{ \eta_1 \eta_2 \eta_3 }
           \frac{ \eta_1 - \eta_2 - \eta_3 }{ 2 }
         + \cdots.
         \label{Ntotal3d}
\end{eqnarray}
Expanding Eq. ({\ref{Ntotal3d})
with respect to $\phi = - \log z$ gives
\begin{eqnarray}
  N &=&
     \frac{ \zeta(3)  }{ \eta_1 \eta_2 \eta_3 }
+    \frac{ \zeta(2)  }{ 2 }
\left(  \frac{ 1 }{ \eta_1 \eta_2 }
      + \frac{ 1 }{ \eta_2 \eta_3 }
      + \frac{ 1 }{ \eta_3 \eta_1 }
 \right)
 - \frac{ \zeta(2) \phi  }{  \eta_1 \eta_2 \eta_3  } \cdots.
     \label{Ntotal3d2}
\end{eqnarray}
As discussed in Section 3.2.1,
$x_1(T_{3D}^{*}) \equiv c_1 = O(1)$ holds at $T = T_{3D}^{*}$.
Inserting this into Eq. (\ref{Ntotal3d2}) gives
\begin{eqnarray}
  N =
     \frac{ \zeta(3)  }{ \eta_1 \eta_2 \eta_3 }
+    \frac{ \zeta(2)  }{ 2 }
 \left[ \frac{ 1 }{ \eta_1 \eta_2 }
 + \frac{ 1 }{ \eta_2 \eta_3 }
   \left( \frac{ 1 }{ 2 } - c_1 \right)
 + \frac{ 1 }{ \eta_3 \eta_1 }
 \right]
    + \cdots
    ~~{\mbox at} ~~T = T_{3D}^{*}.
     \label{Ntotal3d4}
\end{eqnarray}
Thus we have
\begin{eqnarray}
 \frac{ T_{3D}^{*} }{ T_{3D} } = 1 +
 \frac{ c_1 - 1/2 }{ 3 } \frac{ \zeta(2) }{ \zeta(3)^{~2/3} }
 \left( \frac{ k_2 k_3 }{ N }  \right)^{1/3}.
     \label{Ntotal3d5}
\end{eqnarray}
%

%%%%%%%%%%%%%%%%%%%%%%%
%
%   FSE & Crossover T2D*
%
%%%%%%%%%%%%%%%%%%%%%%%

The crossover temperature $T_{2D}^{*}$ associated with two-dimensional
BEC can be defined
if $T_{2D}^{*} < \omega_1$.\footnote{
Thus the system behaves effectively two-dimensional
(EIRD=2) at $T=T_{2D}^{*}$.}
Then from Eq. (\ref{Ntotal3d})
we obtain
\begin{eqnarray}
  N  &=&
     \frac{ g_2( e^{- \eta_2 ( 1 / 2 + c_2 ) } ) }{ \eta_2 \eta_3 }
        + \frac{ g_1( e^{- \eta_2 ( 1 / 2 + c_2 ) } ) }{ \eta_3 }
        +
         \frac{ e^{- \eta_1 / 2 } }{ \eta_1 \eta_2 \eta_3 }
   \nonumber  \\    &+&
     \frac{ \log( 1 - e^{- \eta_1 / 2}  ) \eta_1 }{ 24 \eta_2 \eta_3}
     + \cdots
     \label{Ntotal2d}
\end{eqnarray}
at $T = T_{2D}^{*}$.
When $\eta_1$ becomes large, only the first two terms dominate and we obtain
the crossover temperature as
\begin{eqnarray}
    N &=& \frac{ \zeta(2) }{ \eta_2 \eta_3 }
    - \frac{ 1/2 + c_2 }{ \eta_3 }
    + \frac{ (1/2 + c_2) \log [\eta_2 (1/2 + c_2)] }{ \eta_3 }
    - \frac{ \log (\eta_2 c_2) }{ \eta_3 } + \cdots
%       ~~{\mbox at} ~~T = T_{2D}^{*}
    \nonumber  \\
 &=&
    \frac{ T_{2D}^{*2} \zeta(2) }{ \omega_2 \omega_3 }
  -
    \frac{ T_{2D}^{*} }{ \omega_3 }
    \left[
     (1/2 + c_2) \right.
\nonumber  \\
     &+&
    \left.
  (1/2 + c_2) \log \left( \frac{ T_{2D}^{*} }{\omega_2 (1/2 + c_2)} \right)
     + \log \left( \frac{ T_{2D}^{*} }{\omega_2 c_2} \right)
    \right] + \cdots
    \label{T2dcvr}
\end{eqnarray}
for $\eta_2, \eta_3 << 1$.
For large $N$ and $c_2=1$, we simply have
\begin{eqnarray}
 \frac{ T_{2D}^{*} }{ T_{2D} }
 &=&
  1 +
 \frac{ 5 }{ 8 }
 \left( \frac{ \kappa }{ N \zeta(2) }  \right)^{1/2}
\log
     \left( \frac{ N }{ \kappa \zeta(2) } \right).
     \label{Ntotal3d52}
\end{eqnarray}
%
%%%%%%%%%%%%%%%%%%%%%%%
%
%   FSE & Crossover T1D*
%
%%%%%%%%%%%%%%%%%%%%%%%
One-dimensional crossover temperature $T_{1D}^{*}$
has the same form as in the prolate shape potential case.

The conditions (A) and (B) in Eq. (\ref{f18}) give
the constraints for anisotropy parameters for three-step behavior
to be observed:
\begin{eqnarray}
   & \kappa  &>
         \frac{ N \zeta(2) }{ (\log(2N))^2 },
%   \mbox{ ~( $ T_{1D}^{(0)} < T_{2D}^{(0)} $) },
    \nonumber  \\
   & k_3     &>
                  \frac{ \zeta(3)}{ \zeta(2) }
              \frac{ N }{ \log(2N) }.
%   \mbox{ ~~~~~~~~( $ T_{2D}^{(0)} < T_{3D}^{(0)} $) },
        \label{3step}
\end{eqnarray}
If $T_{2D} << T_{3D}$ holds, since it also implies $T_{2D} << \omega_1$,
excitations along the first axis will be dynamically suppressed
at $T < \omega_1$
making the system effectively two-dimensional before BEC sets in.
Furthermore, if $T_{1D} << T_{2D}$ is also satisfied,
BEC occurs in two-steps, one at $T_{2D}$
and the other at $T_{1D}$.
In Figure 6, above senario of three-step dimensional reduction
is numerically realized.
Three-dimensionally excited modes dominant in higher temperature
are dynamically suppressed at $T < \omega_1$
followed by two-step BEC.
When the condensation into the ground state sets in,
the effective dimension of the system is still two.
From the form of critical temperatures in Eqs. (\ref{Ntotal5}) and (\ref{Ntotal2d52}),
we see that high anisotropy and the small number              
of atoms have similar consequences. 

%%%%%%%%%%%%%%%%%%%%%%%%%%%%%%%%%%%%%%%%%%%%%%%%%%%%%%%%%%%%%%
% begin  add Oct.27,99
%%%%%%%%%%%%%%%%%%%%%%%%%%%%%%%%%%%%%%%%%%%%%%%%%%%%%%%%%%%%%%%
It is useful to see how the weak interaction effect modifies the
above arguments.\footnote { We recover ordinary units in this
discussion for quantitative comparison with other literatures. }
The shift of the critical temperature due to the interaction
effect is given in \cite{GPS96} as $\Delta T_{3D}^{int}/T_{3D}
\sim a N^{1/6} / L$,
where $a$ is the s-wave scattering length, $L \equiv \sqrt{ \hbar
/ m \Omega}$. The shift due to finite size is given in
Eq.(\ref{Ntotal3d5}) as $\Delta T_{3D}^{*}/T_{3D} \sim \left( k_3
/ N \right)^{1/3}$. For a reasonable choice of parameters,
$N=5000$, $k_3=10$, $a/L=0.001$, we obtain $\Delta
T_{3D}^{int}/T_{3D} \sim 0.003$ and $\Delta T_{3D}^{*}/T_{3D} \sim
0.13$. Thus the interaction effect is negligible compared to the
finite size correction.

Comparing the zero point energy in the most confining direction
with the interaction energy, if $\omega_1 > n_0 U$, where $n_0$ is
the ground state density and $U \equiv 4 \pi \hbar^2 a / m$, the
criteria for the system to behave as effectively two-dimensional
is still given by $\omega_1 > T$. For the choice of value $n_0 U /
k_B \sim 110$ [nK] for a sodium atom, this gives  $\omega_1 > n_0
U \sim 10^4$ [1$/$sec]. We should note that the collision effect
even in such a case cannot be strictly two-dimensional
\cite{KSS87}; for high energy atoms, the interaction vertex has
the form of the three-dimensional one. Nevertheless, these effects
will be still suppressed for $a << L_1$ and a small number of
particles whereas the finite size correction behaves as $\Delta
T_{2D}^{*}/T_{2D} \sim \left( \kappa / N \right)^{1/2}$ and
$\Delta T_{1D}^{*}/T_{1D} \sim 1 / \log N$. Therefore the effect
of interaction to the particle number and the critical temperature
remains small and the system is still expected to show the
multistep behavior under these conditions. At $T < \omega_1$, the
most notable difference would be the absence of long range order
due to interactions in this effectively two-dimensional system. In
this case, the system can show quasi-condensation, the
condensation with the fluctuating phase at near $T_{2D}$. The true
condensation with the constant phase will be acheived at lower
temperature. One example of quasi two-dimensional system, a gas of
spin-polarized hydrogen in liquid helium, is known to exhibit
Kosterlitz-Thouless transition \cite{Silvera95,SVYLJ98}. 
Manifestation of
the crossover from BEC to Kosterlitz-Thouless transition
\cite{PHS99} during the multistep process is of particular
interest to study. But even in the presence of such a transition,
we expect that the particle occupation number which is insensitive
to the phase information still behaves similar to the ideal gas.
The interaction contributions used above are calculated within the
framework of the mean field theory. Thus we conclude, apart from
the critical regime where the mean field theory fails,
interactions do not essentially alter the multistep process for a
small number of atoms in highly anisotropic traps.

%%%%%%%%%%%%%%%%%%%%%%%%%%%%%%%%%%%%%%%%%%%%%%%%%%%%%%%%%%%%%%
% end  add Oct.27,99
%%%%%%%%%%%%%%%%%%%%%%%%%%%%%%%%%%%%%%%%%%%%%%%%%%%%%%%%%%%%%%%

From the arguments given in this section,
the sum of most relevant terms in Eq. (\ref{Ntotal3d})
around each crossover temperature will give
the simple expression of the total number of atoms as
%Assuming $1 >> \eta_1 >> \eta_2 >> \eta_3$
%we can simply write (\ref{Ntotal3d}) as
%
\begin{eqnarray}
  N  &=&
        N_0 + N_1 + N_2 + N_3
 \nonumber  \\
  &\sim&
         g_0( e^{- \phi} )
         + \frac{ g_1( e^{- \phi}  ) }{ \eta_3 }
         + \frac{ g_2( e^{- \phi}  ) }{ \eta_2 \eta_3 }
         + \frac{ g_3( e^{- \phi}  ) }{ \eta_1 \eta_2 \eta_3 }.
         \label{Sim1}
\end{eqnarray}
%
%where $\eta_i$ dependent terms in the exponent was absorbed in
%the definition of the constants $c_i$.
Making use of the fact that $\phi$ varies as a nonvanishing function of
the temperature
for the finite system,
we define $N_3(\lambda)
\equiv  g_3( e^{\lambda \phi}  ) / \eta_1 \eta_2 \eta_3 $.
The relation for the Bose function
$\partial_{\phi}  g_n( e^{- \phi} ) = - g_{n-1}( e^{- \phi} )$ \cite{Pathria}
allows us to write
\begin{eqnarray}
  N_2(\lambda)   &\equiv&
 \frac{ g_2( e^{\lambda \phi}  ) }{ \eta_2 \eta_3 }
 = \frac{1}{x_1} \frac{d N_3(\lambda)}{d \lambda},
\nonumber  \\
  N_1(\lambda)   &\equiv&
 \frac{ g_1( e^{\lambda \phi}  ) }{ \eta_3 }
 = \frac{1}{x_1 x_2} \frac{d^2 N_3(\lambda)}{d \lambda^2},
\nonumber  \\
  N_0(\lambda)   &\equiv&
  g_0( e^{\lambda \phi}  )
 = \frac{1}{x_1 x_2 x_3} \frac{d^3 N_3(\lambda)}{d \lambda^3}.
         \label{Sim2}
\end{eqnarray}
Thus the total number of atoms is given by
\begin{eqnarray}
  N  &=&
\frac{1}{x_1 x_2 x_3} \frac{d^3 N_3(-1)}{d \lambda^3}
+ \frac{1}{x_1 x_2} \frac{d^2 N_3(-1)}{d \lambda^2}
+ \frac{1}{x_1} \frac{d N_3(-1)}{d \lambda}
+ N_3(-1) .
         \label{Sim3}
\end{eqnarray}
At $T \sim T_{3D}^{*}$, $x_1 \sim 1$ and $x_2, x_3 >> 1$, then we
have
\begin{eqnarray}
  N  &\sim&
\frac{d}{d \lambda}
N_3(-1)
+ N_3(-1) .
         \label{Sim4}
\end{eqnarray}
At $T \sim T_{2D}^{*}$, $x_2 \sim 1$, $x_1 << 1$, and $x_3 >> 1$,
we have
\begin{eqnarray}
  N  &\sim&
\frac{d}{d \lambda}
N_2(-1)
+ N_2(-1) .
         \label{Sim5}
\end{eqnarray}
At $T \sim T_{1D}^{*}$, $x_3 \sim 1$ and $x_1, x_2 << 1$,
we have
\begin{eqnarray}
  N  &\sim&
\frac{d}{d \lambda}
N_1(-1)
+ N_1(-1) .
         \label{Sim6}
\end{eqnarray}
These results suggest that each condensation fraction $N_i$
behaves similarly as a function of the temperature in the
neiborhood of its characteristic temperature $T_{iD}^{*}$.
%
%Each dimensional component $N_i (i=1,2,3)$ behaves similarly as a
%function of $T /  T_{iD}^{*}$.
We simply write this fact as $N_i(T) \sim F(T / T_{iD}^{*})$ for
$i=1,2,3$, where $F$ is a function independent of $i$. It also
implies that each component has a similar shape in the logarithmic
$T$ scale. The occupation number of each dimension is plotted in
the logarithmic $T$ scale in Figure 7. Note that this derivation
relies on the special property of the Bose function which
determines the density of states of an ideal gas trapped in the
harmonic potential. Whether the same result holds or not in other
systems with different density of states is not obvious.

In conclusion, finite size effects on the Bose-Einstein
condensation of an ideal gas in a strongly anisotropic trap give
rise to various different types of multistep behavior depending on
the degree of anisotropy.
In an isotropic trap, BEC into the ground state always begins
while the system is effectively three-dimensional, i.e. $T_{3D} >
\omega_1=\omega_2=\omega_3$. In an anisotropic trap, in addition
to the BEC which may occur in multisteps, EIRD will also decrease
in steps as the temperature is lowered. The combined effect of
these leads to the appearant multistep behavior. The existence of
the intermediate condensation into one-dimensional space can be
traced back to the logarithmic divergence of the one-dimensional
occupation number in Eq. (\ref{f101a}). This means that, when the
trap is loosened in one direction, the particles tend to occupy
quantum states along this direction with more likelihood than along other
directions. Thus one-dimensionally excited modes in this direction
will dominate multidimensional excitations spread in other
directions ($N_1 >> N_2, N_3$) and the thermodynamic behavior of
such a system is characterized by $T_{1D}$ even though effective
dimension of the system is still three. Note that the same
mechanism is responsible for the {\it nonexistence} of BEC in
one-dimensional harmonic trap in the ordinary thermodynamic limit.
For the same reason, the intermediate condensation into
two-dimensionally excited modes can be observed in a rectangular
cavity \cite{Sonin69,DruKet97}, where $T_{2D}$ does not exist in
the naive thermodynamic limit. Three-step BEC can take place only
in such a system.
Away from the thermodynamic limit,
the temperature dependence of the chemical potential around $T_{1D}$,
$T_{2D}$, and $T_{3D}$ causes similar crossover
behaviors in condensation fractions $N_1$, $N_2$, and $N_3$ as a
function of the reduced temperature.

Atom trap experiments probing the two-step BEC
are realizable in
Ioffe-Pritchard type magnetic traps or
in optical dipole traps \cite{DruKet97}.
The similar type of device
can be used to study multistep behavior discussed here, although
it is difficult to achieve BEC in a maximally anisotropic trap
with our current cooling technique.
Further progress in a trapping device may be required.
The basic mechanism of multistep dimensional crossovers discussed here
can be applied to many other bosonic systems and should be amenable to
future experimental verification.
Quasi low-dimensional systems realized in the optical lattice or waveguide
are the promising option for testing such processes \cite{GHSSPM98,VKCC99}.
Also of interest is
the kinetics of multistep behavior where
the correlation length and the thermal de Broglie wave length are
related in a nontrivial manner.
This is currently under investigation.

%%%%%%%%%%%%%%%%%%%%%%%%%%%%%%%%%%%%%%%%%%%%%%%%%%%%%%%%%%
%
%    ACKNOWLEDGEMENTS
%
%%%%%%%%%%%%%%%%%%%%%%%%%%%%%%%%%%%%%%%%%%%%%%%%%%%%%%%%%%%

\noindent {\bf Acknowledgement}
The author appreciated Prof. B. Hu
for the hospitality of the Center for Nonlinear Studies
at Hong Kong Baptist University
during his visit from March to September 1998.
He also thanks Prof. B. L. Hu for various discussions,
Prof. J. Weiner for useful comments,
particularly of experimental relevance,
and Dr. K. Kirsten for useful references.

%%%%%%%%%%%%%%%%%%%%%%%%%%%%%%%%%%%%%%%%%%%%%%%%%%%%%%%%%%
%
%    APPENDIX
%
%%%%%%%%%%%%%%%%%%%%%%%%%%%%%%%%%%%%%%%%%%%%%%%%%%%%%%%%%%%
\appendix
\renewcommand{\theequation}{\thesection\arabic{equation}}

\section{Dynamical Symmetry in Anisotropic Harmonic Oscillator}
\label{app:DS}
\setcounter{equation}{0}

It is known that dynamical symmetry of anisotropic harmonic oscillators has %%@
the reducible representation
which can be written as the multiple of irreducible
representation of $SU(3)$ symmetry, the %%@
symmetry of isotropic
harmonic oscillators.

For a given $\lambda = (\lambda_1, \lambda_2, \lambda_3) $,
we can define a set of many boson annihilators \cite{BraGre68} as
\begin{equation}
    A_{i}^{\lambda} = a_{i}^{k_i}
    \sqrt{
    [ \frac{ \hat{n}_i }{ k_i } ]
      \frac{ (\hat{n}_i - k_i) !  }{ \hat{n}_i !}
    }
        \mbox{ for $ i = 1,2,3 $ }   ,
      \label{f3}
\end{equation}
where $ \hat{n}_i =  a^{\dag}_i a_i$ are boson number operators.
Many boson annihilators and creators satisfy the following
commutation relations
\begin{equation}
    \left[     A_{i}^{\lambda},   A^{\lambda~\dag}_{j}      \right]
= \delta_{i j}.
      \label{f300}
\end{equation}
Rewriting the Hamiltonian (\ref{f0}) as
\begin{equation}
    H =
    \frac{ \omega }{ 2 }
    \sum_{i=1}^{3}
    \{     A_{i}^{\lambda},   A^{\lambda~\dag}_{i}      \}
  -
    \frac{ 1 }{ 2 }
    \sum_{i=1}^{3}
    \omega_i ( k_i - 2 \lambda_i -1 )       ,
      \label{f4}
\end{equation}
the corresponding energy eigenvalue becomes
\begin{equation}
    E_n =
    \omega (  M + \frac{ 3 }{ 2 } )
    -
    \frac{ 1 }{ 2 }
    \sum_{i=1}^{3}
    \omega_i ( k_i - 2 \lambda_i -1 ).
      \label{c3f5}
\end{equation}
This gives the alternative method to derive energy eigenvalues
given in Eq. (\ref{f2}).

Thus the reducible representation (\ref{f0}) of the original
Hamiltonian leads to the cluster of isotropic harmonic oscillators.
With this decomposition, it is possible to
understand the mechanism of
condensation in anisotropic traps in terms of
condensation in isotropic traps
 \cite{NazDob92}.

\section{Mathematical Formulas}
\label{app:Cumulative Density of States}
\setcounter{equation}{0}

By taking derivatives of
\begin{eqnarray}
 \sum_{M=0}^{\infty} e^{- M \eta}
  = \frac{ 1 }{ 1 - e^{- \eta} }
        \label{App1}
\end{eqnarray}
by $\eta$ on both sides, we obtain
\begin{eqnarray}
 \sum_{M=1}^{\infty} M e^{- M \eta}
  &=& \frac{ e^{- \eta}}{ (1 - e^{- \eta})^{2} }
        \label{App2}
\end{eqnarray}
and
\begin{eqnarray}
 \sum_{M=1}^{\infty} M^2 e^{- M \eta}
  &=& \frac{ e^{- \eta} (1 + e^{- \eta}) }{ (1 - e^{- \eta})^{3} } .
        \label{App21}
\end{eqnarray}
For a small $\eta$, we have an expansion
\begin{eqnarray}
  \frac{ e^{ - \eta } }{ 1 - e^{ - \eta } }
=
  \frac{ e^{ - \eta / 2 } }{ 2 \sinh( \eta / 2 ) }
=
  \frac{ e^{ - \eta / 2 } }{ \eta } - \frac{ \eta e^{ - \eta / 2 } }{ 24 }
    + \cdots.
        \label{App0}
\end{eqnarray}
Similarly, we have
\begin{eqnarray}
 \frac{ e^{- k \eta}}
  { ( 1 - e^{- k \eta} ) ( 1 - e^{- \eta} ) }
% &=&
% \frac{ e^{\eta}}
%  { k \eta^2 }
% \frac{ k\eta}
%  { e^{ k \eta } - 1 }
% \frac{ \eta }
%  { e^{\eta} - 1 }
%  \nonumber  \\
  &=&
%
% \frac{ e^{\eta}}
%  { k \eta^2 }
%   \sum_{n=0}^{\infty}
% \frac{ B_p (k\eta)^p }
%  { p! }
%   \sum_{q=0}^{\infty}
% \frac{ B_q \eta^q }
%  { q! }
%  \nonumber  \\
%
  \frac{ e^{- (k-1) \eta / 2 }}
     { k \eta^2 }
 -
 ( k + \frac{1}{k} )
  \frac{ e^{- (k-1) \eta / 2 }}
     { 24 }
    + \cdots
        \label{App3}
\end{eqnarray}
%
%where $B_n$ are the Bernoulli polynomials.
%Their values are given by
%$B_0 = 1 ,B_1 = -1/2, B_2 = 1/6, B_4 = -1/30, \cdots$
%and $B_3 = B_5 = B_7 = \cdots = 0$ \cite{Serre70}.
and
\begin{eqnarray}
 \frac{ e^{- \eta_1}}
  { ( 1 - e^{- \eta_2} ) ( 1 - e^{- \eta_3} ) }
 &=&
 \frac{ e^{-(\eta_1 - \eta_2 - \eta_3) / 2} }
  { 8 \sinh( \eta_1 / 2 ) \sinh( \eta_2 / 2 ) \sinh( \eta_3 / 2 )}
 \nonumber  \\
  =
   \frac{ e^{-(\eta_1 - \eta_2 - \eta_3) / 2} }
  { \eta_1 \eta_2 \eta_3 }
 &-&  \frac{ e^{-(\eta_1 - \eta_2 - \eta_3) / 2} }{ 24 }
  ( \frac{ \eta_1^{~2} + \eta_2^{~2} + \eta_3^{~2} }
   { \eta_1 \eta_2 \eta_3 }
  )
  + \cdots.
        \label{f1510}
\end{eqnarray}
From Eq. (\ref{App0}),
\begin{eqnarray}
    \sum_{l=1}^{\infty}
    \frac{ z^l e^{ - l \eta } }{ 1 - e^{ - l \eta } }
&\sim&
    \sum_{l=1}^{\infty}
    \frac{ z^l e^{ - l \eta / 2 } }{ l \eta }
    =
  - \frac{1}{ \eta }
  \log( 1 - z e^{- \eta / 2} ).
    \label{App100}
\end{eqnarray}
%

%%%%%%%%%%%%%%%%%%%%%%%%%%%%%%%%%%%%%%%%%%%%%%%%%%%%%%%%%%
%
%    BIBLIOGRAPHY
%
%%%%%%%%%%%%%%%%%%%%%%%%%%%%%%%%%%%%%%%%%%%%%%%%%%%%%%%%%%%

%
%\vspace{1cm}
%
%
\begin{flushleft}
{\large\bf Figure Captions \\}
%\listoffigures
%\begin{figure}
\end{flushleft}

\noindent {\bf Figure 1}
The condensation temperatures $T_{1D}$ (solid curve),
 $T_{2D}$ (dashed curve), and
 $T_{3D}$ (dot-dashed curve) are shown
as a function of $k_3$. $N=10^4$,
$\omega_1=0.5$, $k_2$ is also varied as $k_2^{~2}=k_3$.
The logarithmic scale is used for both axis.

\vspace{0.5cm}

\noindent {\bf Figure 2}
Different condensation behaviors corresponding to different
anisotropy parameters
$\omega_1 / \omega_2$ and $\omega_2 / \omega_3$ are indicated.
$N=10^4$ is chosen. The logarithmic scale is used for both
horizontal and vertical axis.
$\zeta(2)$ and $\zeta(3)$ are approximated as one for simplicity.

\vspace{0.5cm}

\noindent {\bf Figure 3}
The condensation fractions
$N_0 / N$ (solid curve),  $N_1 / N$ (dashed curve),
$N_2 / N$ (dot-dashed curve),  $N_3 / N$ (dotted curve)
as a function of the
temperature for an isotropic trap are plotted.
The same symbols are used in Figures 4-7.
$\omega_1=0.1$, $\omega_2=0.1$, $\omega_3=0.1$, and $N=1000$
are chosen.
$T_c = 0.94$ is the three-dimensional critical temperature
in Eq. (\ref{T3d2}).

\vspace{0.5cm}

\noindent {\bf Figure 4}
The condensation fractions for a prolate shape anisotropic trap are plotted.
$\omega_1=0.3$, $\omega_2=0.3$, $\omega_3=0.0003$, and $N=10^4$
are chosen.
Two-step condensation occurs in this case.
$T_c = T_{3D}^{*} = 0.61$ from Eq. (\ref{Ntotal5}) and
$T_{1D}^{*} = 0.34$ from Eq. (\ref{T1dcvr2}).
$c_1$ and $c_3$ are set equal to one.
Note that EIRD=3 during the whole two-step BEC process.

\vspace{0.5cm}

\noindent {\bf Figure 5}
The condensation fractions for an oblate shape anisotropic trap are plotted.
$\omega_1=0.3$, $\omega_2=0.002$, $\omega_3=0.002$, and $N=10^3$
are chosen.
%
%$T_{3D}^{*} = 0.17$ is the three-dimensional crossover temperature
%in Eq. (\ref{Ntotal2d5}).
$T_c = T_{2D}^{*} = 0.057$ is the two-dimensional crossover temperature
in Eq. (\ref{Ntotal2d52}).
$c_2$ is set equal to one.
The system shows the dimensional reduction from three to two dimension
at $T = \omega_1 = 0.3$.
Hence the accumulation of particles into $N_2$ from $N_3$
is not the result of condensation.
At $T < \omega_1$, the system show the ordinary two-dimensional
condensation into the ground state.

\vspace{0.5cm}

\noindent {\bf Figure 6}
The condensation fractions for a maximally anisotropic trap are plotted.
Three-step dimensional reduction can be seen.
$\omega_1=0.3$, $\omega_2=0.02$, $\omega_3=0.0004$, and $N=5 \times 10^3$ are used.
$T_{3D}^{*} = 0.17$, $T_{2D}^{*} = 0.06$, and $T_{1D}^{*} = 0.02$
are crossover temperatures defined
in Eqs. (\ref{Ntotal3d5}), (\ref{Ntotal3d52}), and (\ref{T1dcvr2}),
respectively.
$c_1=c_2=c_3=1$ are used.
The system shows the dimensional reduction
 at $T = \omega_1 = 0.3$ similar to Figure 5,
and behaves two-dimensionally near $T < 0.3$.
Therefore, the accumulation of particles into $N_2$ from $N_3$
is not the result of condensation.
For $T < \omega_1$, the system shows two-step BEC
at $T_{2D}^{*}$ into one-dimensionally excited states and
at $T_{1D}^{*}$ into the ground state.
%
%$\omega_1=0.5$, $\omega_2=0.04$, $\omega_3=0.0002$, and $N=600$ are used.
%$T_{3D}^{*} = 0.24$, $T_{2D}^{*} = 0.06$, and $T_{1D}^{*} = 0.02$
%

\vspace{0.5cm}

\noindent {\bf Figure 7}
The logarithmic $T$ scale is used for the three-step behavior
shown in Figure 6.
The same parameters as in Figure 6 are used.

\end{document}